\documentclass{ws-ijmpcs}
\begin{document}

\markboth{M.~Albaladejo, C.~Hidalgo-Duque, J.~Nieves, and E.~Oset}
{Hidden charm molecules in a finite volume}

%
\catchline{}{}{}{}{}
%


\title{HIDDEN CHARM MOLECULES IN A FINITE VOLUME}

\author{M.~ALBALADEJO}
\author{C.~HIDALGO-DUQUE}
\author{J.~NIEVES}
\author{E.~OSET}
\address{Departamento de F\'{\i}sica Te\'orica and IFIC, Centro Mixto Universidad de 
Valencia-CSIC\\
Institutos de Investigaci\'on de Paterna, Aptdo. 22085, 46071 Valencia,
Spain}
\maketitle

\begin{history}
\received{Day Month Year}
\revised{Day Month Year}
\end{history}

\begin{abstract}
In the present paper we address the interaction of charmed mesons in hidden
charm channels in a finite box. We use the interaction from a recent model based on heavy quark
spin symmetry that predicts molecules of hidden charm in the infinite volume. The energy
levels in the box are generated within this model, and several methods for the analysis of these levels (``inverse problem'') are investigated.

\end{abstract}

\ccode{PACS numbers: 11.25.Hf, 123.1K CHANGE}

\section{Introduction}
In this paper we report about our recent work on hidden charm molecules in finite volume.\cite{Albaladejo:2013aka} Lattice QCD (LQCD) is becoming an increasingly powerful tool to study the hadron spectrum from a theoretical point of view. One of the tools becoming gradually more used is the analysis of LQCD levels in terms of the L\"uscher method.\cite{Luscher:1985dn} This method converts binding energies of a hadron-hadron system in the finite box into phase shifts of the hadron-hadron interaction from levels above threshold, or binding energies
from levels below threshold. Recently, this method has been generalized\cite{Doring:2011vk} by using the on-shell factorization scheme used in the unitary chiral approach.\cite{aoorev}

The purpose of this paper is to study the possibility of finding heavy--meson molecules, predicted long time ago,\cite{Voloshin:1976ap} in future LQCD simulations. To study these states, we shall follow the formalism of Ref.~\refcite{HidalgoDuque:2012pq}, where an effective field theory incorporating light SU(3)-flavor and heavy-quark spin symmetries to describe charmed meson-antimeson
interactions is formulated, generalizing the work of Ref.~\refcite{Nieves:2012tt}. In studying such molecules in finite volume our objective is twofold. First, by putting the model in a box, we make predictions about the interacting energy levels that could be tested in actual LQCD simulations, as that of the work in Ref.~\refcite{Prelovsek:2013cra}. Second, we will propose several methods to face the problem of analyzing the energy levels predicted with the model (``inverse problem''), as it would be done if they were real LQCD results.

\section{Formalism}
To study the interactions between charmed mesons, we use the formalism derived in Refs.~\refcite{HidalgoDuque:2012pq,Nieves:2012tt}. The (non-relativistic) $T$-matrix for the charmed mesons scattering is written as:
\begin{equation}\label{eq:tmat}
T^{-1}(E) = -\frac{\mu k}{2\pi} \cot\delta - i \frac{\mu k}{2\pi} = V^{-1}(E) - G(E)~.
\end{equation}
In this equation, $\mu$ is the reduced mass of the two-meson system, $k$ is the momentum, $E$ is the energy, $\delta$ is the phase shift, $G$ is a one-loop function, and $V$ is the interaction potential. The loop function $G$ provides the right-hand or unitarity cut (RHC), and the contributions of the left-hand cut (LHC), if any, are included in the potential $V$. The loop function $G$ is regularized by means of a Gaussian form factor with a cutoff $\Lambda$, and it is given by:
\begin{align}
G(E) & = 
\int \frac{\text{d}^3 \vec{q}}{(2\pi)^3} \frac{e^{-2(\vec{q}^{\,2}-k^2)/\Lambda^2}}{E-m_1-m_2 - \vec{q}^{\,\,2}/2\mu + i0^+} \nonumber\\
& = -\frac{\mu\Lambda}{(2\pi)^{3/2}}e^{2k^2/\Lambda^2} + \frac{\mu k}{\pi^{3/2}} \phi\left(\sqrt{2}k/\Lambda\right)-i \frac{\mu k}{2\pi}~, \qquad \phi(x) = \int_{0}^{x} e^{y^2} \text{d}y~.\label{eq:gmat_gr}
\end{align}

In the approach of Refs.~\refcite{HidalgoDuque:2012pq,Nieves:2012tt}, the leading-order (LO) interaction comes from contact terms, and the potential in Eq.~\ref{eq:tmat} is written as:
\begin{equation}\label{eq:pot}
V(E) = e^{-2k^2/\Lambda^2} C(\Lambda)~,
\end{equation}
where $C$ is a constant. HQSS and light-flavour SU(3) symmetry reduce the number of independent constants to only four so that, for each channel, the constant $C$ is the appropriate linear combination of these four independent constants. For these constants, we take the values given in Ref.~\refcite{HidalgoDuque:2012pq}. A more complete description of the formalism can be found in Ref.~\refcite{Albaladejo:2013aka}. In particular, there they are discussed the absence of any contribution to the LHC in the LO potential, and the role of the cutoff $\Lambda$.

The amplitude in a box of size $L$, denoted by $\widetilde{T}$, is written from Eq.~\eqref{eq:tmat} by replacing the integral of the loop function $G$ with a discrete sum over the allowed momenta $\vec{q} = \frac{2\pi}{L}\vec{n}~,\ \vec{n} \in \mathbb{Z}^3$ (periodic boundary conditions), giving rise to a new function, $\widetilde{G}$:
\begin{equation}
\widetilde{T}^{-1}(E) = V^{-1}(E) - \widetilde{G}(E)~, \qquad
\widetilde{G}(E) = \frac{1}{L^3}\sum_{\vec{q}} \frac{e^{-2(\vec{q}^{\,2}-k^2)/\Lambda^2}}{E-m_1-m_2 - \vec{q}^{\,2}/2\mu}.
\end{equation}
Now, the energy levels in the box are given by the poles of the
$\widetilde{T}$-matrix, $V^{-1} = \widetilde{G}$. For the energies of these levels, the amplitude in the infinite volume is recovered as:
\begin{equation}
\label{eq:euluscher}
T^{-1}(E) = V^{-1}(E)-G(E) = \widetilde{G}(E) - G(E) \equiv \delta G(E)~.
\end{equation}
Further discussions about the finite volume aspects of the formalism are to be found in Ref.~\refcite{Albaladejo:2013aka}. In particular, there it is delivered an analytic treatment of the dependence of the $\delta G(E)$ function on the cutoff $\Lambda$ and its relation with the L\"uscher function ${\cal Z}_{00}(1,\hat{k}^2 )$. Additionally, there can also be found a discussion about the  absence of the LHC in our approach and the the finite volume effects the latter could produce.

\section{Results}

\begin{figure}
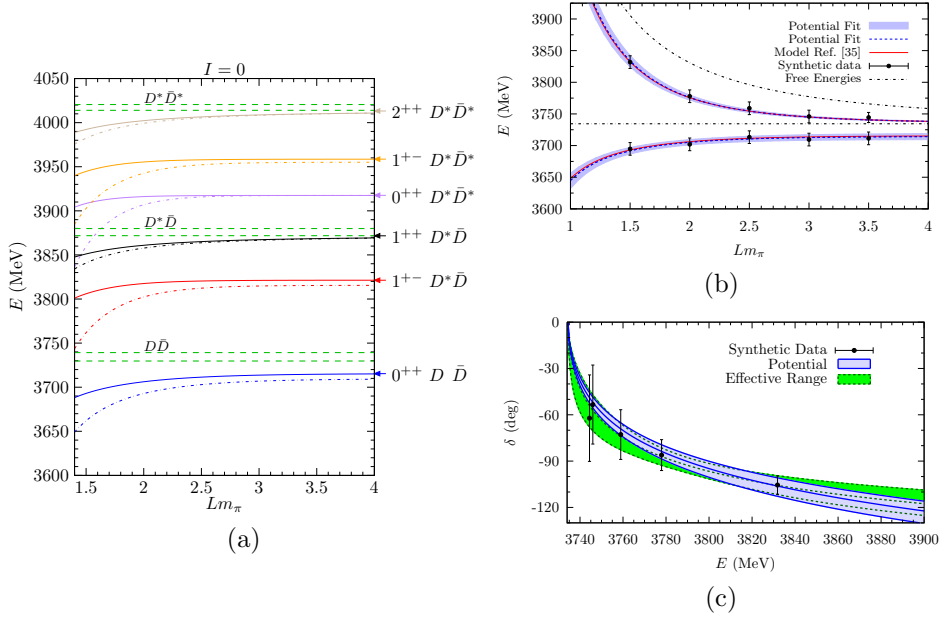
\centering%
\begin{minipage}[c]{0.5\textwidth}%
\includegraphics[height=6.0cm,keepaspectratio]{BoundStatesFigures_0.mps}\\%
\centering(a)%
\end{minipage}%
\begin{minipage}[c]{0.5\textwidth}%
\includegraphics[height=3.4cm,keepaspectratio]{FigurasProblemaInverso_1.mps}\\%
\centering(b)\\%
\vspace{2mm}
\includegraphics[height=3.4cm,keepaspectratio]{FigurasProblemaInverso_0.mps}\\%
\centering(c)%
\end{minipage}%
\caption{(a) Some subthreshold energy levels for different channels. (b) The first two energy levels for the $D\bar{D}$ $I=0$ $J^{PC}=0^{++}$ interaction. (c) Phase shifts for this interaction.\label{fig:pi}}
\end{figure}

In Fig.~\ref{fig:pi}(a) we show the energy levels obtained from the poles of the $\widetilde{T}$-matrix as a function of $L$. We only show those levels that correspond to bound states in the infinite volume case. An attractive interaction in a finite volume generates subthreshold energy levels. In the infinite volume case, these energy levels can asymptotically tend to threshold or, instead, become bound states. As said before, the finite volume energy levels found here could be compared with those obtained in real LQCD simulations. In an actual LQCD simulation, however, one would obtain a finite (and typically, small) number of points scattered on Fig.~\ref{fig:pi}(a), rather than the theoretical continuous curves shown. Then, some of the levels shown in Fig.~\ref{fig:pi}(a) could be safely identified with bound states, asymptotically different from threshold, but other energy levels could not. Hence, it becomes necessary to analyze the energy levels in some manner, as we do next. 

In Fig.~\ref{fig:pi}(b), we show the first two energy levels (below and above threshold) for the case of the $I=0$ $J^{PC}=0^{++}$ $D\bar{D}$ interaction. For this channel, the infinite volume model predicts a bound state with mass $E_B=3715\ \text{MeV}$. The energy levels in the box, predicted by the model, are given by the red solid lines. The subthreshold level tends for $L\to\infty$ to the bound state mass. From the theoretical levels, we generate {\it synthetic data} (see Ref.~\refcite{Albaladejo:2013aka} for details), shown by the points with its errors. We analyze these points as if they were actual LQCD data (inverse analysis). The first method of analysis consists in obtaining the phase shifts from the upper level of Fig.~\ref{fig:pi}(b), with the standard L\"uscher's formula. The phase shift so obtained for each point of the upper level is shown with the points in Fig.~\ref{fig:pi}(c). From the phase shifts, then, one calculates the effective range parameters. The green band in Fig.~\ref{fig:pi}(c) represents the phase shift calculated with these parameters and the effective range expansion for $k\ \cot\delta$. From these parameters, in turn, the mass of the bound state (if any) can be calculated. With this method, we obtain $E_B=3721^{+10}_{-25}\ \text{MeV}$. A second method of analysis consists in parameterizing the potential in Eq.~\eqref{eq:pot}, and fit the parameters to reproduce the energy levels. The energy levels of the best fit are shown in Fig.~\ref{fig:pi}(b) by the blue dashed line and its associated error band. With this potential, then, one can go back to the infinite volume model, and obtain the mass of the bound state, which turns out to be $E_B=3715^{+3}_{-6}\ \text{MeV}$. In Ref.~\refcite{Albaladejo:2013aka} the independence of the method with respect to the regulator chosen is also shown. A third method is also employed, by parameterizing the amplitude in the infinite volume with an effective range expansion, and then analyzing both levels simultaneously. It is important to note that the difference with respect to the first method is two-fold: the energy levels, and not the phases, are analyzed, and both levels (above and below threshold) instead of only one (below threshold) are taken into account. From the effective range parameters so obtained, one gets $E_B=3716^{+4}_{-5}\ \text{MeV}$. The second and third methods, as opposed to the first one, yield central values for the mass in better agreement with the one of the infinite volume model, and, furthermore, the errors so determined are smaller. We conclude that these methods turn out to be more efficient than the first one, and that they can help in analyzing energy levels obtained in real LQCD simulations.


\end{document}